\documentclass[conference, 10pt]{IEEEtran}

\usepackage{cite}
\usepackage{url}
\usepackage{graphicx}
\usepackage{color}
\usepackage{placeins}
\usepackage{float}
\usepackage{tabularx,colortbl}
\usepackage{ifthen}
\usepackage{xspace}

\usepackage{tipa} 

\usepackage[breaklinks, plainpages=false, colorlinks=true, anchorcolor=blue, citecolor=blue, linkcolor=blue, urlcolor=blue, bookmarks=false]{hyperref}

\renewcommand{\thefootnote}{\fnsymbol{footnote}}
\makeatletter
\newcommand\blfootnote[1]{%
	\begingroup
	\renewcommand{\@makefntext}[1]{\noindent\makebox[1.8em][r]#1}
	\renewcommand\thefootnote{}\footnote{#1}%
	\addtocounter{footnote}{-1}%
	\endgroup
}
\makeatother

\makeatletter

\newcounter{author}
\renewcommand{\author}[2][]{
   \stepcounter{author}
   \@namedef{author@\theauthor}{#2}
   \@namedef{authorlabel@\theauthor}{#1}
}

\newcounter{address}
\newcommand{\address}[2][]{
   \stepcounter{address}
   \@namedef{address@\theaddress}{#2}
   \@namedef{addresslabel@\theaddress}{#1}
}

\newcommand{\alsep}{and}

\def\newmaketitle{\par%
  \begingroup%
  \normalfont%
  \def\thefootnote{}
  \def\footnotemark{}
  \let\@makefnmark\relax
  \footnotesize
  \footnotesep 0.7\baselineskip
  \normalsize%
  \twocolumn[\thenewmaketitle\@IEEEaftertitletext]%
  \if@IEEEusingpubid
     \enlargethispage{-\@IEEEpubidpullup}%
  \fi
  \endgroup
  \setcounter{footnote}{0}\let\maketitle\relax\let\@maketitle\relax
  \gdef\@thanks{}%
  \let\thanks\relax}

\def\thenewmaketitle{
  \newpage
  \begin{center}%
    \vskip0.2em{\Huge\@IEEEcompsoconly{\sffamily}\@IEEEcompsocconfonly{\normalfont\normalsize\vskip 2\@IEEEnormalsizeunitybaselineskip
   \bfseries\large}\@title\par}\vskip1.0em\par%
    \vspace{1ex}
    \newcounter{c@author}
    \newcounter{c@tmp}
    \ifthenelse{\value{author}=2}{%
      \newcommand{\liand}{ and }}{%
      \newcommand{\liand}{, and }}
    \ifthenelse{\value{address}<2}{%
      \@nameuse{author@1}%
      \stepcounter{c@author}%
      \whiledo{\value{c@author}<\value{author}}{%
        \setcounter{c@tmp}{\value{author}}%
        \addtocounter{c@tmp}{-\value{c@author}}%
        \ifthenelse{\value{c@tmp}=1}{%
          \renewcommand{\alsep}{\liand}}{\renewcommand{\alsep}{, }}%
        \stepcounter{c@author}\alsep \@nameuse{author@\thec@author}}\\%
    }
    {
      \@nameuse{author@1}${}^{(\ref{\@nameuse{authorlabel@1}})}$%
      \stepcounter{c@author}%
      \whiledo{\value{c@author}<\value{author}}{%
      \setcounter{c@tmp}{\value{author}}%
      \addtocounter{c@tmp}{-\value{c@author}}%
      \ifthenelse{\value{c@tmp}=1}{%
        \renewcommand{\alsep}{\liand}}{\renewcommand{\alsep}{, }}%
      \stepcounter{c@author}\alsep \@nameuse{author@\thec@author}%
        ${}^{(\ref{\@nameuse{authorlabel@\thec@author}})}$%
      }
    }
    \vspace{0.2ex}

    \ifthenelse{\value{address}>0}{%
      \ifthenelse{\value{address}=1}{
        {\@nameuse{address@1}}
      }
      {
        \newcounter{c@address}

        \begin{center}
        \whiledo{\value{c@address}<\value{address}}
        {
          \refstepcounter{c@address}
            ${}^{(\thec@address)}$\,%
              \label{\@nameuse{addresslabel@\thec@address}}%
              \@nameuse{address@\thec@address}\\ %
        }
        \end{center}
      } 
    }
    {
      \relax
    }
  \end{center}
}

\makeatother

\newcommand{\kko}{k'ni\textipa{P}atn k'l$\left._\mathrm{\smile}\right.$stk'masqt\xspace}

\newcommand\aj{AJ} 
\newcommand\apj{ApJ} 
\newcommand\apjs{ApJS} 
\title{\huge A VLBI Calibration System with Real-time Pulsar Gating for \\ FRB Localization using CHIME/FRB Outriggers}

\author{Aaron~B.~Pearlman\textsuperscript{(1,2,*)} for The CHIME/FRB Collaboration \vspace{0.1cm}}

\newcommand{\MCGILL}{Department of Physics, McGill University, 3600 rue University, Montr\'eal, QC H3A 2T8, Canada}
\newcommand{\TSI}{Trottier Space Institute, McGill University, 3550 rue University, Montr\'eal, QC H3A 2A7, Canada \newline (*)~Corresponding author: \href{mailto:aaron.b.pearlman@physics.mcgill.ca}{aaron.b.pearlman@physics.mcgill.ca}}

\address{(1)~\MCGILL \break (2)~\TSI \footnote[1]{text}}

\begin{document}

\newmaketitle


\begin{abstract}

Several thousand fast radio burst~(FRB) sources have been discovered using the Canadian Hydrogen Intensity Mapping Experiment~(CHIME) radio telescope, as part of the CHIME/FRB project. Currently, CHIME/FRB is able to localize most FRBs to a limiting precision of several arcminutes, which can be improved to subarcminute precision for some FRB sources through offline analysis of their baseband data. This allows only the most nearby sources to be robustly associated with a host galaxy. Using three new Outrigger telescopes located at transcontinental distances from CHIME, the CHIME/FRB Outriggers project will improve the localization capabilities of CHIME/FRB. Together, these radio telescopes will form a wide field of view, very long baseline interferometry~(VLBI) array that will enable FRBs discovered by CHIME/FRB to be localized to a limiting precision of $\sim$50 milliarcseconds. The astrometric position of each FRB will be determined using calibration solutions derived from well-localized radio pulsars and compact, steady radio sources. We present an overview of the VLBI calibration system that will be employed within the CHIME/FRB Outriggers project to achieve high precision FRB localizations, which will enable studies of a large number of FRB host galaxies and local environments.

\end{abstract}


\section{Introduction}

The three CHIME/FRB Outrigger telescopes each consist of a single cylindrical reflector that is oriented to observe the same field of view as CHIME, and they operate in the same radio frequency range (400--800\,MHz) as CHIME/FRB~\cite{CHIME+2018}. The Outrigger telescope at \kko Observatory~(KKO) is located in Penticton, British Columbia (66\,km from CHIME) and is equipped with 64 dual-polarization antennas, with 1/16$\textsuperscript{th}$~($\sim$500\,m\textsuperscript{2}) of CHIME's collecting area. The other two Outrigger telescopes are located at Green Bank Observatory~(GBO) in Green Bank, West Virginia (3300\,km from CHIME) and at Hat Creek Radio Observatory~(HCRO) in Hat Creek, California (1000\,km from CHIME). The Outrigger telescopes at GBO and HCRO have 1/8$\textsuperscript{th}$~($\sim$1000\,m\textsuperscript{2}) of CHIME's collecting area and are outfitted with 128 dual-polarization antennas. When an FRB is detected by CHIME/FRB, a trigger is sent to each Outrigger telescope to save the buffered baseband voltage data containing the FRB signal for cross-correlation and localization using synoptic, transient VLBI techniques~\cite{Leung+2021, Cassanelli+2022, Cassanelli+2023}.

Traditionally, VLBI has been performed on known targets. The CHIME/FRB Outriggers project will instead perform VLBI on FRB targets that are mostly not known a priori and can produce coherent radio emission anytime or anywhere on the sky. Since this VLBI program is being carried out in tandem with CHIME/FRB's blind, wide field of view real-time FRB search, we are developing a novel VLBI calibration system for constructing calibration solutions, primarily using pulsars as astrometric calibrators, that will be used to determine positions for FRBs detected using CHIME/FRB. Here, we provide an overview of the VLBI calibration system that will be used as part of the CHIME/FRB Outriggers project.


\section{VLBI Calibration System}

\begin{figure*}
	\hspace{-1.1cm}
	\includegraphics[trim=0cm 0cm 0cm 0cm, clip=true, scale=0.20, angle=0]{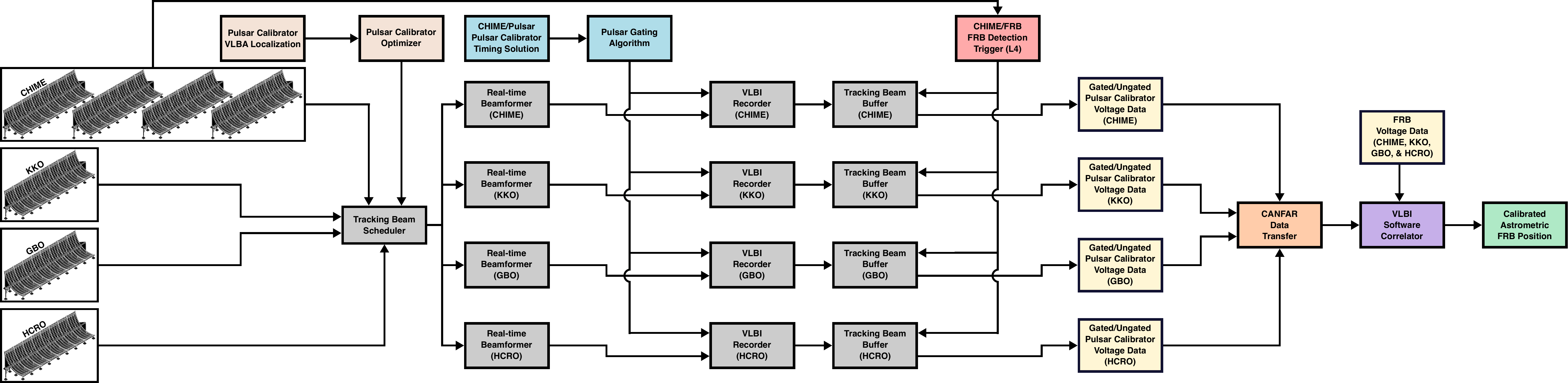}
	\caption{Overview of the VLBI calibration system for CHIME/FRB Outriggers.}
	\vspace{-0.4cm}
	\label{fig:f1}
\end{figure*}

In Figure~\ref{fig:f1}, we show a schematic diagram of the main components of the VLBI calibration system. Brief descriptions of the main system components are provided below.


\subsection{Pulsar Calibrators}

\begin{figure}
	\centering
	\includegraphics[trim=0cm 0.25cm 0cm 0.2cm, clip=true, scale=0.346, angle=0]{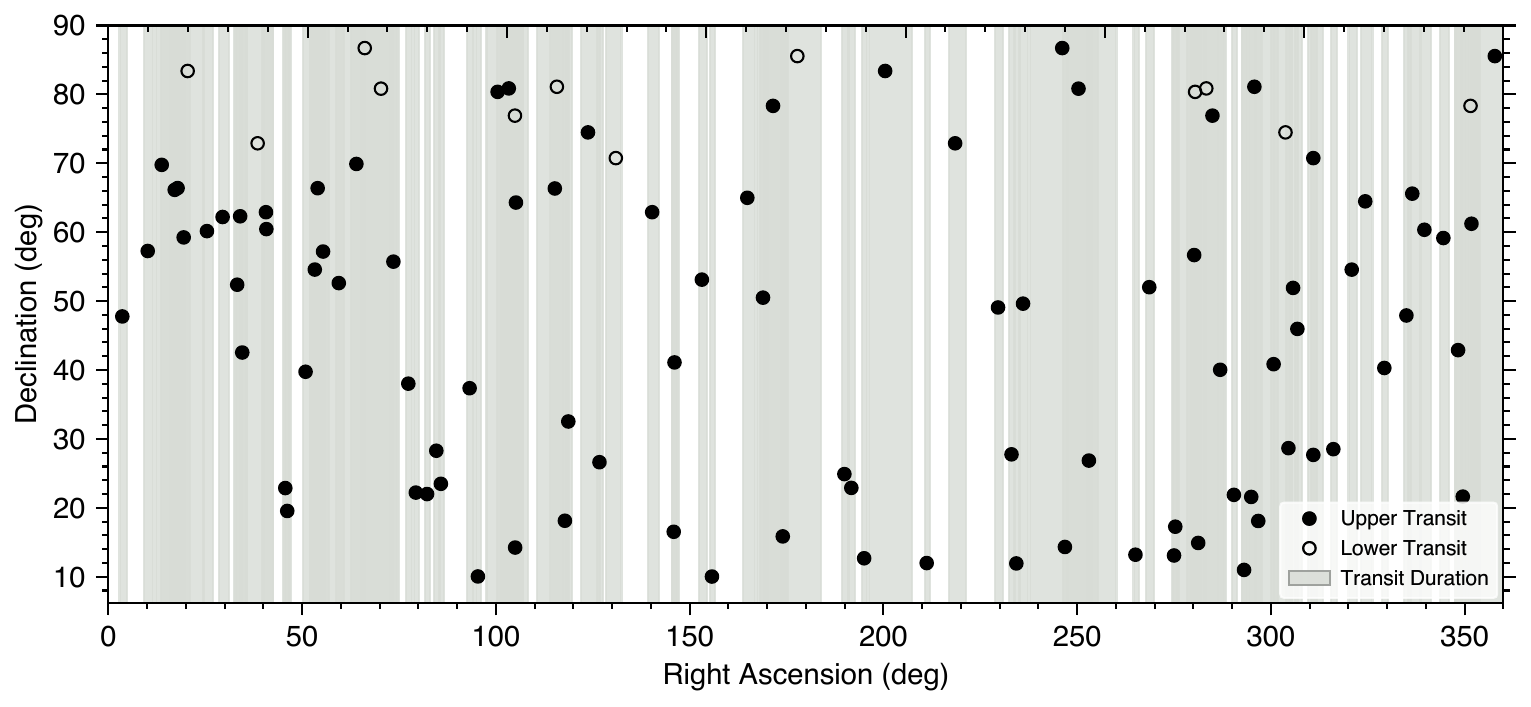}
	\vspace{-0.6cm}
	\caption{Radio pulsars used for VLBI calibration as part of the CHIME/FRB Outriggers project. The filled and open black circles show the sky position of each pulsar calibrator during their upper and lower transits, respectively, through the primary beam of CHIME. The gray shaded regions indicate the transit durations of each pulsar.}
	\vspace{-0.5cm}
	\label{fig:f2}
\end{figure}

A sample of $\sim$100 radio pulsars, distributed across CHIME/FRB's field of view, will be used as astrometric calibrators for the CHIME/FRB Outriggers project (see Figure~\ref{fig:f2}). The pulsars in this sample were selected based on their brightness, timing properties, and sky position. A calibration solution will be derived for each FRB from observations of pulsar calibrators performed close in time and near each FRB target, and carried out simultaneously with CHIME/FRB and the Outrigger telescopes. VLBI observations with the Very Long Baseline Array~(VLBA) have been performed to precisely measure the positions and proper motions of each pulsar calibrator (VLBA/21A-314, PI:~Kaczmarek; VLBA/22A-345, PI:~Curtin; VLBA/23A-099, PI:~Curtin), in order to facilitate FRB localizations with precisions down to $\sim$50 milliarcseconds using CHIME/FRB and the Outrigger telescopes.


\subsection{Pulsar Timing}

The CHIME/Pulsar system~\cite{CHIME+2021a} is used to perform high cadence radio observations of each pulsar calibrator using two of its ten tracking beams. These observations are used to derive pulsar timing ephemerides, which are regularly updated and checked for timing anomalies, such as glitches. The timing solutions of each pulsar calibrator are used for determining the times when the voltage data from the ``on-pulse'' region of each pulsar calibrator should be saved at each telescope site.


\subsection{Pulsar Gating}

Prior to each pulsar calibrator observation, the CHIME/Pulsar timing solution of each calibrator is used to generate a polynomial that precisely predicts the rotational phase of the pulsar at all times during upcoming observations at each telescope site. The polynomial ephemeris is constructed using the \texttt{tempo} pulsar timing software package. We precompute the times of the ``on-pulse'' region for each pulsar at each recorded radio frequency using the pulsar's dispersion measure~(DM). This information is transmitted to the VLBI backend (FX correlator) at each telescope site, which allows the ``on-pulse'' voltage data at each frequency to be saved in real-time for offline cross-correlation analyses. This technique is referred to as \textit{pulsar gating}.


\subsection{VLBI Backend at CHIME and the CHIME/FRB Outriggers}

At each Outrigger site, the \texttt{kotekan} data processing framework~\cite{Renard+2021} is run on multiple FX correlator nodes to produce two tracking beams using a real-time beamforming kernel. The two tracking beams at each Outrigger telescope, along with two tracking beams from the CHIME/Pulsar system, are used to perform simultaneous observations of the pulsar calibrators during their transits through the primary beam of CHIME. The observations are scheduled to allow raw voltage data to be recorded from each pulsar calibrator simultaneously at each telescope site. Channelized, dual-polarization gated voltage data, with (4+4) bit complex numbers per sample, are recorded across the 400--800\,MHz band at a time resolution and frequency resolution of 2.56\,$\mu$s and 0.39\,MHz, respectively. The gated voltage data from the Outrigger telescopes are initially stored in a large ring buffer ($\sim$1\,TB at KKO and $\sim$2\,TB at GBO and HCRO) in RAM on the X-engine nodes. The gated pulsar calibrator data are saved to disk at the Outrigger sites when an FRB trigger is received from CHIME/FRB. The gated voltage data from the two CHIME/Pulsar tracking beams are saved directly to 2\,$\times$\,6\,TB high-endurance Non-Volatile Memory Express~(NVMe) solid state drives~(SSDs), which serve as the buffer at the CHIME site. After the data are saved to disk, the FRB and pulsar calibrator data are transferred to a large computer cluster at the Canadian Advanced Network for Astronomical Research~(CANFAR), where the data are cross-correlated and an astrometric position is measured for each FRB.

The VLBI calibration system is designed to provide gated voltage data from up to two nearby pulsar calibrators before and after each FRB. The VLBI recorders can also be operated in non-gating mode, which provides flexibility to observe compact, steady radio sources for calibration purposes as well. The VLBI calibration system is under rapid development, and it will significantly aid in the removal of instrumental, atmospheric, and ionospheric effects when the visibilities are phase referenced after the data are cross-correlated. The calibration solutions provided by this system will be essential for achieving the design FRB localization precision of $\sim$50 milliarcseconds using synoptic, transient VLBI techniques.


\section*{ACKNOWLEDGEMENT}

A.B.P. is a Banting Fellow, a McGill Space Institute~(MSI) Fellow, and a Fonds de Recherche du Quebec -- Nature et Technologies~(FRQNT) postdoctoral fellow.




%

\end{document}